\documentclass[journal]{IEEEtran}

\usepackage{cite}
\usepackage{array}
\usepackage{graphicx}
\usepackage{dcolumn}
\usepackage[utf8]{inputenc}
\usepackage[T1]{fontenc}
\usepackage{amsmath}
\usepackage{amssymb}

\begin{document}
%
\title{On the Effect of Low Temperatures on the Maximum Output Power of a Coherent Erbium Doped Fiber Amplifier}
%
%
%

\author{Julien Le Gouët, Jérémy Oudin, Philippe Perault, Alaeddine Abbes, Alice Odier, and Alizée Dubois}%
\thanks{Julien Le Gouët, Jérémy Oudin, Philippe Perault are with Office National d'Etudes et de Recherches Aérospatiales (ONERA), Palaiseau, France, e-mail: julien.le\_gouet@onera.fr }
\thanks{Alaeddine Abbes was with ONERA and is now at IES, Univ Montpellier, UMR CNRS 5214, Montpellier, France}
\thanks{Alice Odier was with ONERA and is now at Amplitude Systemes, Pessac, France}
\thanks{Alizée Dubois was with ONERA and is now at Service de Physique de l'Etat Condensé, CEA, Gif/Yvette, France}

\maketitle

\begin{abstract}
The influence of low temperatures on the performance of a high-power single-frequency fiber laser amplifier is evaluated with a numerical simulation. Cooling the fiber can allow to take advantage of both higher damping of the acoustic waves in the silica glass, and higher laser efficiency. We first report on the measurement of the stimulated Brillouin scattering (SBS) threshold in a silica fiber as a function of the temperature from 300~K down to 77~K. We then present the measurements of small-signal absorption and gain spectra of an erbium doped alumino-phosphosilicate fiber, at room temperature and liquid nitrogen temperature. Based on these data, we derive a numerical study of the combined effects of cooling on the SBS threshold and the amplifier efficiency, and conclude on the interest of this technique for SBS limited high power Er doped fiber amplifiers (EDFA). The temperature increase caused by the pump laser in the fiber core is also addressed.
\end{abstract}


%
\IEEEpeerreviewmaketitle

\section{Introduction}

\IEEEPARstart{A}{mong} the countless types of lasers, high power sources with single frequency and single spatial mode emission have numerous applications. The amplification of coherent pulses (longer than 100 ns) up to the kW peak power range allow the long distance optical measurement of wind velocity \cite{Dolfi_2009}, and continuous wave (cw) sources at 100~W or more \cite{Creeden_2016} are interesting for free-space telecommunications or defense applications. The single spatial mode offers the best atmospheric propagation, while the spectral linewidth defines the velocity measurement resolution or the capacity limit of coherent combining.

For the best compactness and robustness, these systems are preferably based on fiber coupled laser sources, which are amplified in rare-earth doped optical fibers. As it is well-known, for coherent sources with spectral linewidth narrower than 10~MHz, the maximum power in fibers is limited by the stimulated Brillouin scattering (SBS) \cite{Kobyakov10}. In order to increase the power delivered by these fiber amplifiers, several approaches have been developped to hinder the SBS effect. In essence, they come down to either increase the effective mode area \cite{Ahmad2017}, or reduce the interaction length \cite{Shi2010}, or reduce the SBS gain. This last solution can be obtained for example by reducing the overlap between acoustic and optical waves \cite{Li2007}, by introducing a thermal \cite{Imai_1993} or strain gradients \cite{Yoshizawa1993} along the fiber to generate an inhomogeneous broadening of the SBS gain, or by changing the very nature of the core glass \cite{Ballato2013}.

Another degree of freedom is the influence of low temperatures on the attenuation of the acoustic waves in silica glass. This effect has been analyzed since the 50's \cite{Pine69,Anderson55}, but its various physical origins have been understood in more detail only recently \cite{Courtens07}. For hypersonic waves generated by optical radiations at a wavelength of 1.5~$\mu$m, the attenuation presents a maximum at about 100~K. This higher damping of the hypersound reduces the negative impact of the Brillouin backscattering, and allows the propagation of higher power in the fiber.

Cooling the fiber has the other well-known effect of improving the efficiency for lasers based on rare earth doped crystals \cite{Dubinski09} or fibers \cite{Yao2011}. This effect is due to the Boltzmann population distribution in the lower Stark levels, increasing the absorption cross-section in the vicinity of the zero-phonon transition.

The aim of this article is to study the influence of low temperatures on the maximum output power of a typical single-frequency EDFA. This study is based on two measurements between room and liquid nitrogen temperatures: the SBS linewidth and threshold power in a passive fiber, and the spectra of absorption and gain of an erbium (Er) doped single-mode fiber. To our knowledge, this is the first study about the combined effects of SBS linewidth and amplifier efficiency on the maximum reachable power at low temperature.

As a brief state of the art of the cryogenic cooling of optical fibers, we can mention several applications. In passive single-mode fibers, the spectral shift and width of the SBS spectrum are considered to monitor the thermal stability of superconductor magnets cooled down to a few Kelvins, such as those used at the LHC in CERN or in ITER \cite{EPFL02,MIT08}. 
In the case of Er doped fibers, the spectral properties of the laser amplification have been analyzed down to 4~K, in order to analyze the influence of the Er gain homogeneous linewidth in wavelength multiplexing telecommunications  \cite{Desurvire_1990}. Single-mode Er doped fibers are exposed to even lower temperatures in demonstrations of quantum memories compatible with fiber telecommunications \cite{Tittel_2015}. As for high power lasers at cryogenic temperatures, numerous works have been dedicated to Ytterbium doped large mode area (LMA) fibers, either to increase the laser efficiency \cite{Jelger09}, or to improve the gain in the low wavelength domain \cite{Steinborn13}. It should be highlighted that throughout this literature, no particular issue was reported about the resistance of the fibers (core or clad pumped) to the low temperatures. 


In the following, we first present the measurements of the SBS threshold in a passive single mode fiber as a function of the temperature. Then we report our measurements of the small-signal interaction spectra (absorption and emission) on a single-mode Er:Yb doped alumino-phosphosilicate fiber \cite{Bubnov2009}. We use these data as inputs for a numerical simulation of a single frequency high power amplifier (>100~W range) in an Er doped LMA fiber pumped in the same transition as the laser (so-called in-band pumping, here in the $^4I_{15/2}-^4I_{13/2}$ transition). Finally we conclude on the improvement of efficiency and maximum power that can be obtained by cooling the active fibers.

\section{\label{part_II}Temperature dependence of SBS threshold power}

The basic idea of this work stems from a well-referenced variation of the hypersound attenuation in amorphous media as a function of the temperature \cite{Anderson55}. 
In a crystal host, the hypersonic damping is dominated on a large temperature range by the phonon-phonon interactions, which involve the contribution of the thermal bath phonons. In that range (typically 20-300~K), the SBS linewidth (inverse of the hypersound lifetime) varies linearly with the temperature. However in noncrystalline solids, the thermal variation of the Brillouin linewidth is not monotonous \cite{Pine69}. Because of the combination of several mechanisms, this variation exhibits a broad maximum at a temperature $T$ that increases with increasing acoustic frequency, following an Arrhenius law \cite{Hertling_1998, Lefloch01}.

In the case of standard silica optical fibers, the acoustic wave is generated at approximately 11~GHz for a  1.5~$\ \mu$m signal wavelength \cite{Agrawal_book}. For that acoustic frequency, the SBS linewidth $\Delta\nu_B$ has been measured to increase from 50~MHz to 95~MHz when cooling from 300~K to 130~K \cite{Lefloch03}. Across the same temperatures, the Brillouin frequency shift varied by only 2\% \cite{Lefloch01}.

For a single-frequency signal, with a spectral linewidth small compared to the SBS linewidth, the SBS small-signal gain writes as:

\begin{equation}
\label{eq_gB}
g_B=\dfrac{2\pi\gamma_e^2}{nv_{ac}c\lambda_s^2\rho_0\Delta\nu_B}
\end{equation}
where $n$ is the core refractive index, $c$ the velocity of light in vacuum, $\lambda_s$ the wavelength of the optical signal field, $v_{ac}$ the acoustic waves velocity in the glass core, $\gamma_e$ is the electrostrictive coupling constant and $\rho_0$ the mean value of the glass density \cite{Boyd_ONL}. If the ratio $\gamma_e^2/\rho_0$ does not vary significantly with $T$, an increase of $\Delta\nu_B$ translates into a decrease of $g_B$. This finally corresponds to an increase of the incident SBS threshold power, according to its usual expression:

\begin{equation}
\label{eq_seuilSBS}
P_\text{th}=21\dfrac{A_{ao}}{g_B L_\text{eff}}
\end{equation}
where $A_\text{ao}$ is the acousto-optic area \cite{Kobyakov10}, and $L_\text{eff}$ is the interaction effective length, defined as :

\begin{equation}
\label{eq_interaction_Leff}
L_\text{eff}=\dfrac{1}{P_\text{sig}(L)} \int^{L}_{0} P_\text{sig}(z) dz
\end{equation}

\subsection{SBS linewidth measurements}

\begin{figure}[h!]
	\includegraphics[width=9cm, trim={0cm 0 -0cm 0}, clip]{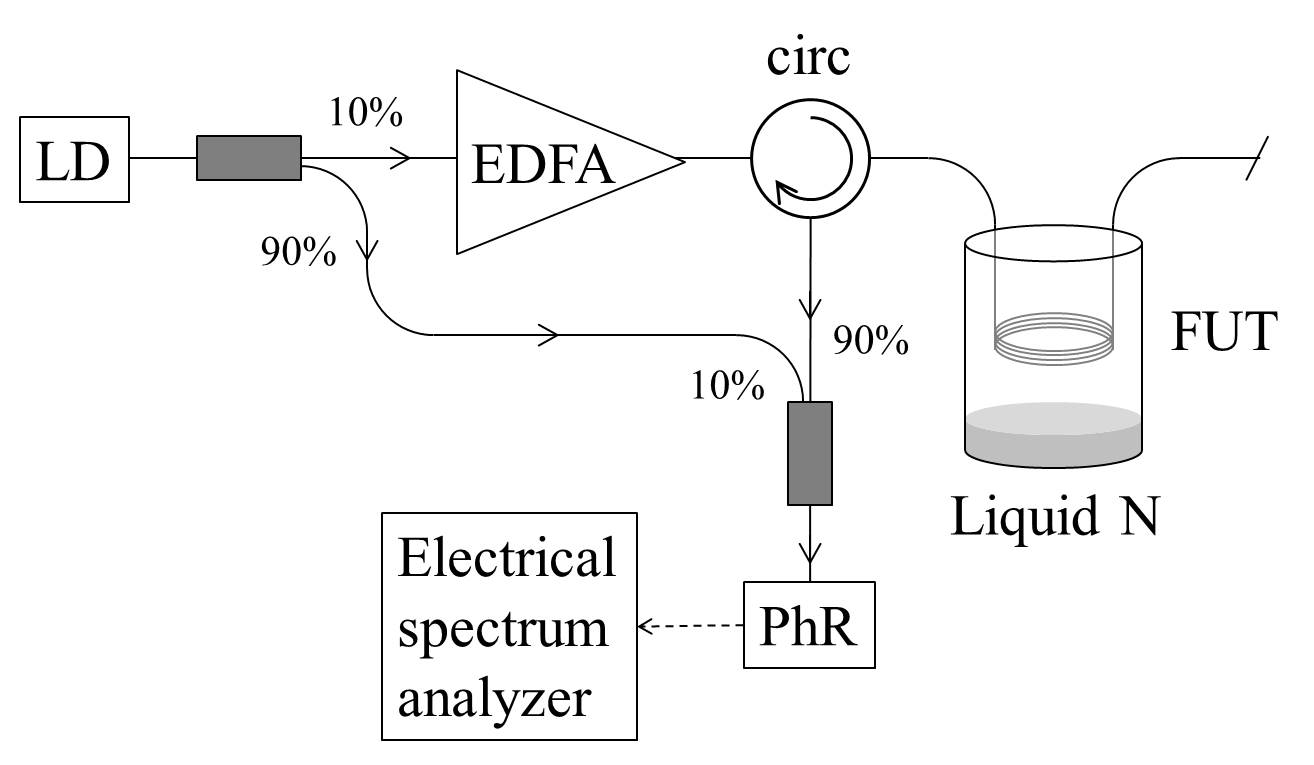}
	\caption{Setup for the measurement of the SBS spectrum linewidth.}
	\label{Fig_Setup_linewidth} 
\end{figure}

We first measure the SBS linewidth as a function of the temperature, as reported in previous works \cite{Lefloch03}, for a 10~m long single-mode PM980 fiber (SMF). The method consists in analyzing the heterodyne beat-note between the injected signal and the SBS generated Stokes wave (Figure \ref{Fig_Setup_linewidth}). For the signal, we use a laser diode (LD) at 1545~nm with linear polarization. The signal is split between a local oscillator and a probe arm. The probe is amplified by an EDFA before injection in the SMF. The power after the EDFA is high enough to provide a measurable SBS spectrum, and low enough to avoid spectral narrowing by the gain in a stimulated regime of Brillouin scattering \cite{Boyd_1990}. The SMF output is cleaved at an angle of $10^{\circ}$ to prevent any reflection that could seed the SBS process. A PM circulator is used to collect the Stokes wave, which has the same polarization as the input signal. The interference with the local oscillator is finally measured on a 12 GHz band photoreceiver, and analyzed on a 26~GHz band electric spectrum analyzer.

The SMF is coiled on a 12~cm diameter aluminum support, and the temperature of the SMF can be varied between 77~K and 300~K by adjusting the height of the coil above a liquid nitrogen bath. Three thermistors are placed on the circumference of the fiber support to check the homogeneity of the SMF temperature. We find that in our nitrogen container, a vertical displacement of 1 cm corresponds to a temperature shift ranging from 5~K close to the top, to 15~K close to the nitrogen bath.

The measurement of the SBS linewidth as a function of $T$ is displayed on Figure \ref{Fig_03_23_linewidth_empirical_def} (gray squares). It shows that the SBS linewidth finds a local maximum of about 75~MHz for $T \approx 100 \ \text{K}$, like in previously reported measurements. At room temperature, the SBS linewidth is about 40~MHz, so we expect that the SBS threshold can be improved by a factor of 1.9.

\subsection{\label{part_II.1}Empirical definition of SBS}

\begin{figure}[h!]
	\includegraphics[width=9cm, trim={4cm 1.5cm -0cm 0}, clip]{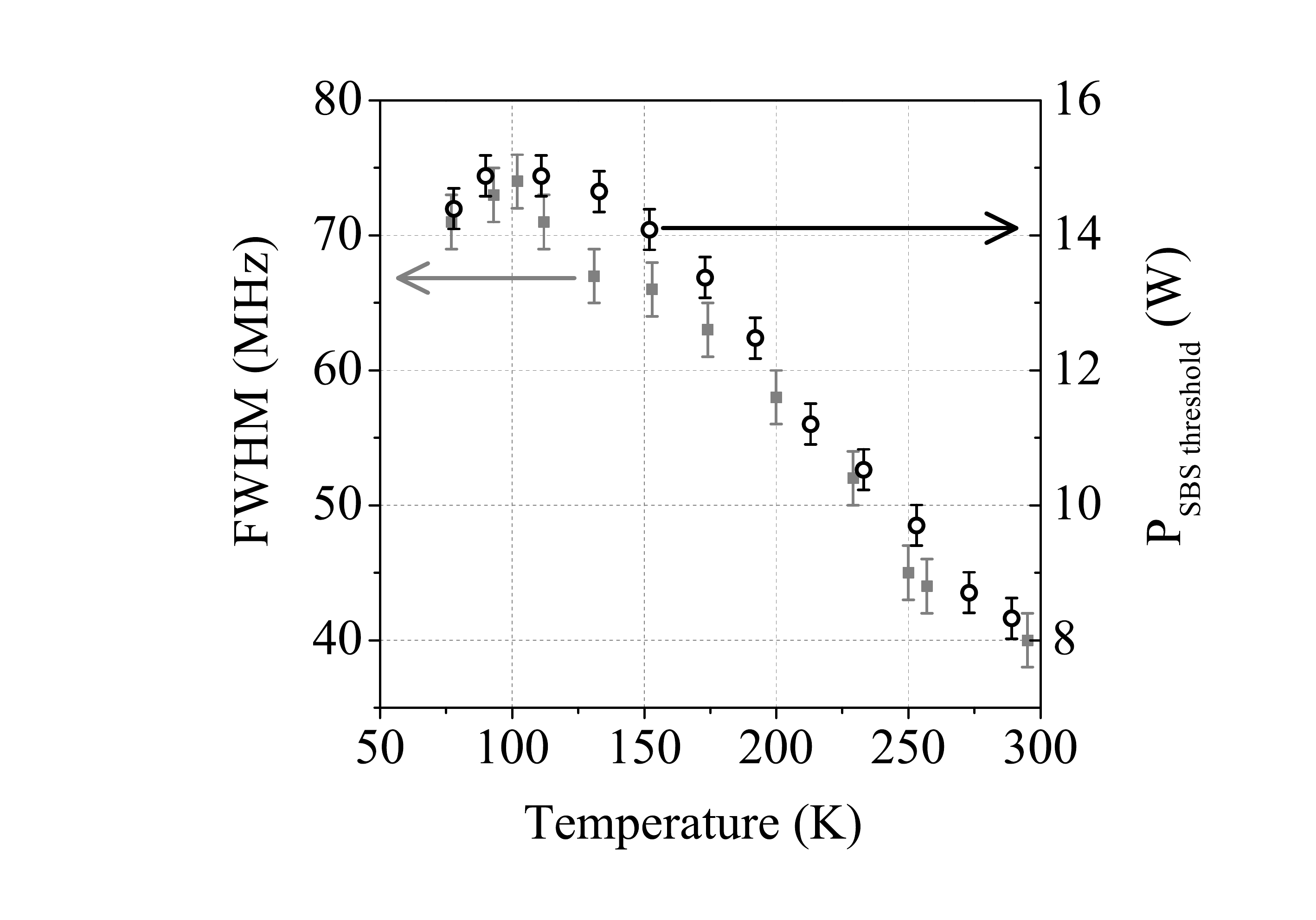}
	\caption{Thermal variations of the SBS spectral linewidth (gray squares) and of the SBS threshold power (black circles), as defined empirically by the alteration of the output pulse shape.} 
	\label{Fig_03_23_linewidth_empirical_def} 
\end{figure}

If the linewidth $\Delta\nu_B$ is the only parameter that varies significantly with temperature in the SBS gain expression (see eq. (\ref{eq_gB})), then the influence of $T$ on the SBS threshold (eq. (\ref{eq_seuilSBS})) can be directly obtained from the variation of $\Delta\nu_B$ (see Figure \ref{Fig_03_23_linewidth_empirical_def}, gray squares). In order to verify this correlation, we proceed to an empirical measurement that consists in monitoring the shape of high power pulses at the fiber output. Once the Brillouin scattering reaches the amplifying regime, close to the SBS threshold, a noticeable fraction of the input signal is backscattered. For a given input pulse shape, this translates into an alteration of the output pulse shape close to the peak.

The method for this measurement is straightforward: an acousto-optic modulator generates laser pulses from the 1545~nm laser diode, with a pulse duration of 400~ns and a repetition rate of typically 20~kHz. A two-stage EDFA can raise the injected peak power at the level of the SBS threshold in the 10~m SMF. The input peak power can be modified finely by tuning the repetition rate. The output pulse shape is monitored on a photodiode with a rise time of 10~ns. We define the empirical SBS threshold by the input power required to generate a noticeable dip of about 10\% of the peak power. The relative error of this measurement is lower than 10\%.

From Figure \ref{Fig_03_23_linewidth_empirical_def} (black circles), we observe that the SBS threshold increases from 8~W at 300~K to 15~W at 100~K, which corresponds to an improvement by a factor of 1.9. Between 300~K and 77~K, the SBS threshold increases by a factor 1.8. We also observe a very good correlation between the SBS threshold and the SBS linewidth, as suggested by eq. (\ref{eq_gB}) and (\ref{eq_seuilSBS}) with the approximation that $\gamma_e^2/\rho_0$ does not vary significantly with T. Therefore this approximation is confirmed with a relative error lower than 10\%.

\subsection{Relative variation of the threshold}

The SBS threshold variation with temperature can be confirmed by measuring the backscattered power as a function of the input power. The setup here is a direct measurement of the Stokes wave power, at the output of the optical circulator in the previous setup (Fig. \ref{Fig_Setup_linewidth}). A direct time-resolved measurement cannot differentiate the contributions of the elastic Rayleigh and inelastic Brillouin backscattered waves, so we use an optical spectrum analyzer (OSA), with a spectral resolution of 0.05~nm. The corresponding frequency resolution of 6~GHz at 1545~nm allows the discrimination of the two contributions, typically separated by 11~GHz in silica fibers. 

\begin{figure}
	\includegraphics[width=10cm, trim={4cm 1.7cm 2cm 0}, clip]{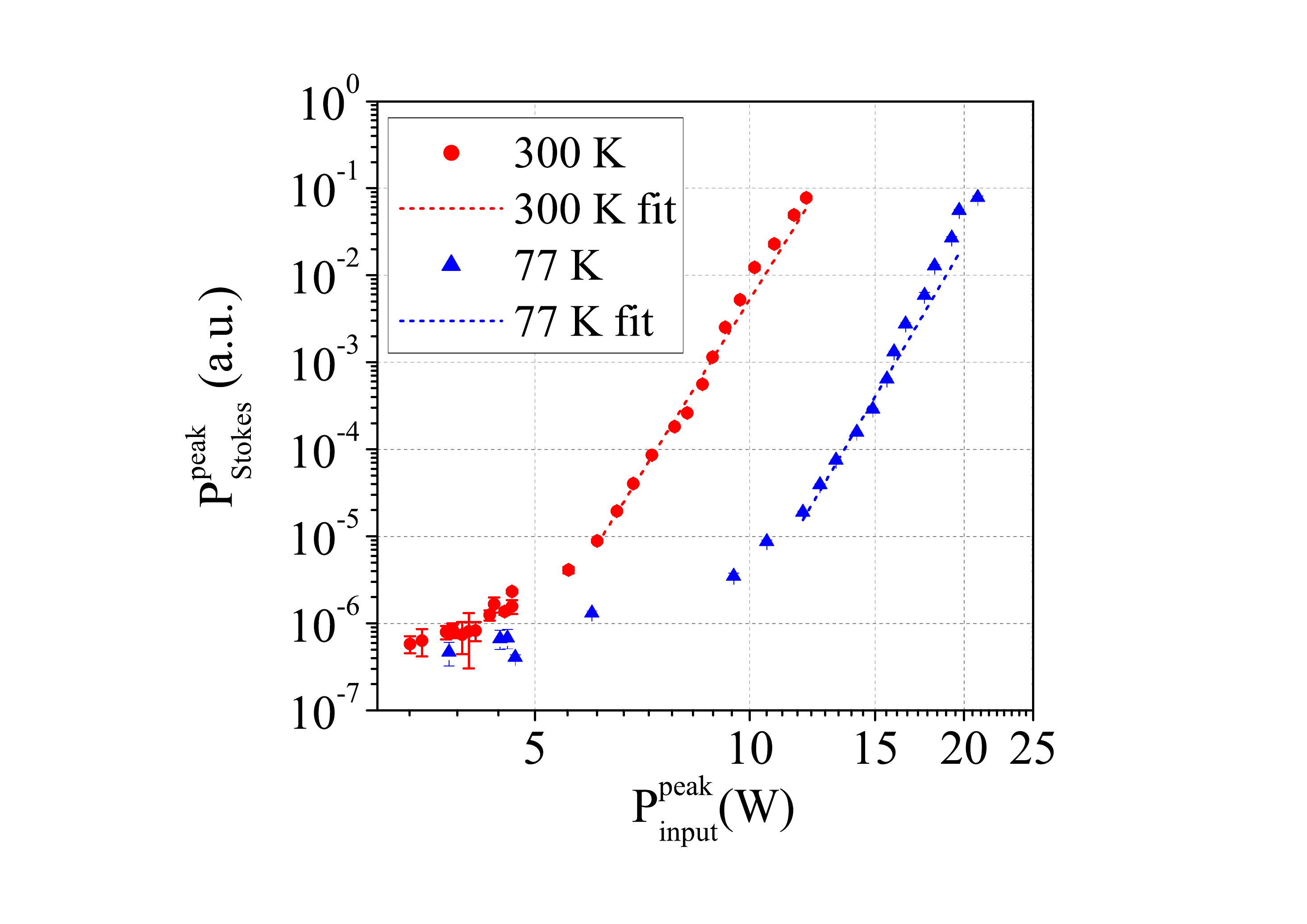}
	\caption{Peak power of the backscattered Stokes wave as a function of the coupled peak power, at 77 (blue triangles) and 300~K (red circles). The dotted lines represent the corresponding exponential fits (blue and red dotted lines) in the regime of stimulated Brillouin scattering with undepleted pump.}
	\label{Fig_03_24_comparison_threshold} 
\end{figure}

On Figure \ref{Fig_03_24_comparison_threshold} the peak power of the SBS Stokes wave is plotted as a function of the input peak power, for the fiber immersed in the liquid nitrogen bath (77~K) and outside (300~K). We did not perform the measurement at intermediary temperatures, because the acquisition of one curve is long enough to let some nitrogen evaporate, which would modify the height of the coil relative to the bath, and finally modify the temperature during the measurement. 

For each curve, two regimes can be identified, corresponding to the spontaneous and undepleted stimulated Brillouin scattering. For 300~K, the spontaneous regime does not appear clearly, as the backscattered power is too close to the noise floor of the OSA. For 77~K it appears as the first section, for $P_\text{input}^\text{peak} < 10 \ $W. The regime of undepleted stimulated Brillouin scattering appears for $P_\text{input}^\text{peak} \gtrsim 5 \ $W at 300~K, and for $P_\text{input}^\text{peak} \gtrsim 10 \ $W at 77~K. In this regime, the curves can be fitted with a power function:
\begin{equation}
\label{eq_fitSBS}
P_\text{Stokes}=a.P_\text{input}^b
\end{equation}
As indicated by the parallelism of the curves at 77 and 300~K, the exponent $b$ is identical for both curves ($b=13.5 \pm 0.5$). For the amplitude coefficients, the fits yield $a_{300}=(1.8 \pm 0.1).10^{-16}$ and $a_{77}=(5.1 \pm 0.4).10^{-20}$.

The SBS threshold power corresponds to the incident power $P_\text{th}$ that induces the backscattering of some arbitrary fraction $\eta_\text{th}=P_\text{Stokes}/P_\text{th}$ of this input power. Hence, the criterion $\eta_\text{th}$ can be chosen differently, for example if the the setup under study is more or less sensitive to the backscattered power. Given the expression (\ref{eq_fitSBS}) used to fit the data, the fraction $\eta_\text{th}$ can be expressed as:
\[\eta_\text{th}=a_\text{300} \cdot P_\text{th\ 300}^{b-1}=a_\text{77} \cdot P_\text{th\ 77}^{b-1}\]
which yields for the ratio of the SBS threshold powers at 77 and 300~K:
\begin{equation}
\label{eq_ratio_thresholds}
\dfrac{P_{\text{th}\ 77}}{P_{\text{th}\ 300}}=\left( \dfrac{a_\text{300}}{a_\text{77}}\right)^{\dfrac{1}{b-1}}=1.8 \pm 0.1
\end{equation}
Given the relation \ref{eq_seuilSBS} between the threshold power and the SBS gain, we deduce that $g_B$ is 80\% lower at 77~K than at room temperature.

The result of this quantitative measurement is identical to the ratio obtained with the empirical determination of the SBS threshold (see Fig. \ref{Fig_03_23_linewidth_empirical_def}). Therefore it confirms the relevance of the first method to evaluate the influence of the temperature on $P_\text{th}$.

\section{Influence of the temperature on the amplification efficiency}

The efficiency of a rare-earth based laser is known to increase at low temperatures, thanks to the redistribution of the excited state populations in the Stark sub-levels that are closer to the zero-phonon line (about $1.53 \ \mu \text{m}$ for Erbium in usual silica glasses). For signals at longer wavelengths, the absorption cross-section is strongly reduced, and the effective electronic structure corresponds to a four-level system \cite{Millar_90,Kagi_91}. We present here our measurements of small-signal absorption and gain spectra at room and liquid nitrogen temperatures on a particular Er:Yb doped alumino-phosphosilicate (Al:P) fiber, and we investigate numerically the interest of cooling the active fiber in a high power single-frequency amplifier.

\subsection{Brief spectroscopy of Er in Yb codoped alumino-phosphosilicate fiber}

The Al:P host has attracted a lot of interest in view of Erbium:Ytterbium codoping in silica fibers \cite{Vienne1996,Unger2007}. When Yb is used to increase the pumping efficiency of Er, a phosphorus doping is required to prevent the energy back-transfer from Er to Yb : the P$_2$O$_5$ glass formers increase the maximum phonon energy and favor the multi-phonon decay, thus reducing the $^4I_{11/2}$ lifetime in Er \cite{Vienne_1998}. On the other hand, Al oxide codoping is required to improve the dilution of $\text{Er}^{3+}$ ions. Otherwise, for high doping concentrations, the ions can form clusters that share the laser pump energy and reduce its efficiency \cite{Myslinski_1997}. By doping silica glass with both Al and P oxides, it was soon discovered that the local formation of $\text{AlPO}_4$ units in the silica network could reduce the refractive index of the core, leading to even lower index than the value of the pure silica for equimolar concentrations \cite{DiGiovanni1989}. This possiblity of tailoring the refractive index profile is particularly interesting to design large mode area (LMA) fibers with single spatial mode. In addition, several groups have reported the interest of adding Yb or Yb:P codopants in fibers with high Er concentrations to help the solubility of $\text{Er}^{3+}$ ions in the silica network \cite{Lim2012,Kiritchenko2015}. Therefore the combination of Yb codoping and Al:P host appears as an interesting candidate for Er doped fibers for high efficiency amplifiers. 

\begin{figure}
	\includegraphics[width=8.5cm, trim={0cm .5cm 0cm 0}, clip]{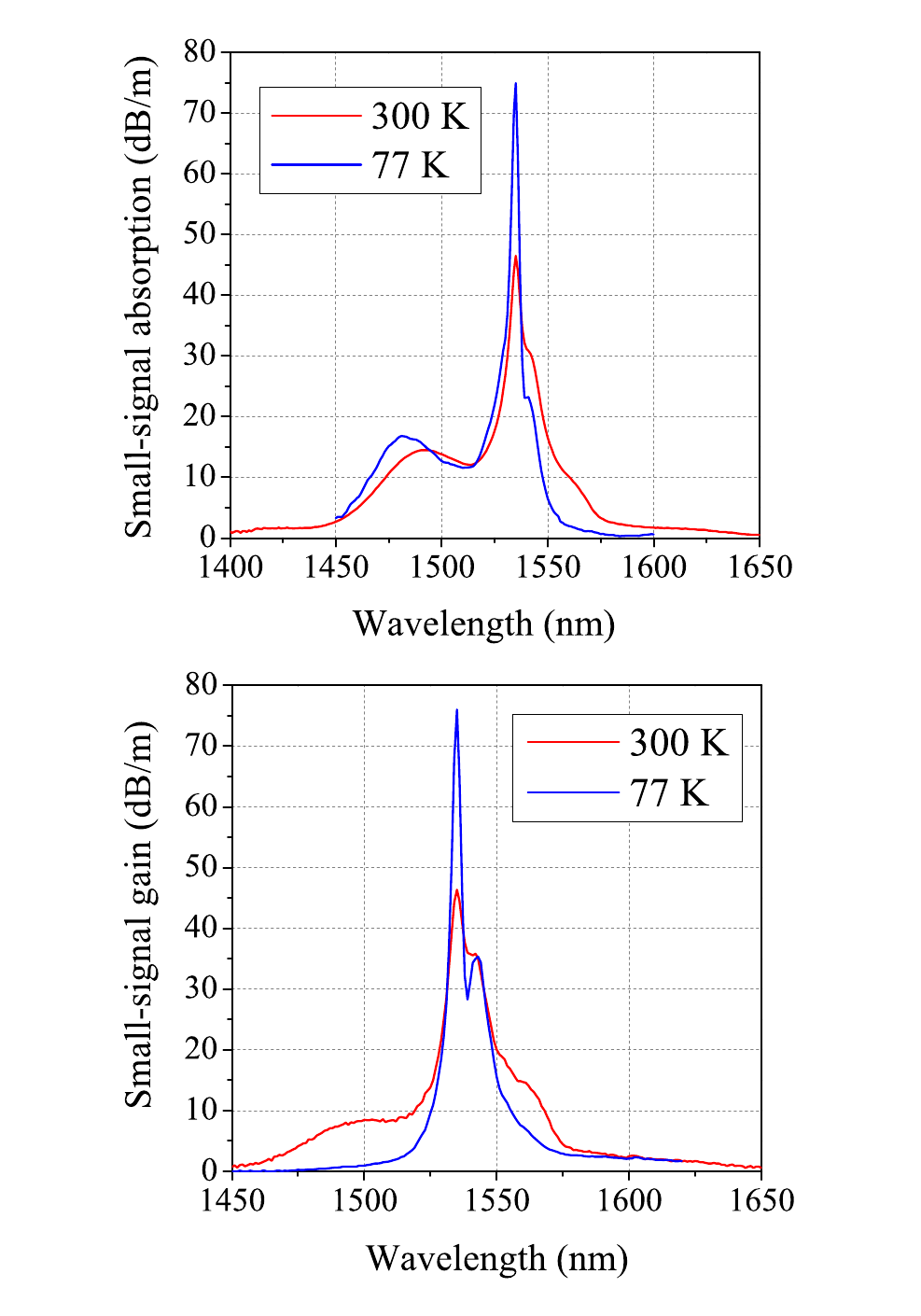}
	\caption{Small-signal absorption (above) and gain (below) spectra, at 300~K (red) and 77~K (blue).}
	\label{Fig_12_05_spectra} 
\end{figure}

As a first confirmation of this guess, we evaluated an important parameter in a higly doped fiber: the fraction $2k$ of Er ions in pairs among the total population. A common method consists in measuring the ratio of non-saturable and small-signal absorptions at the peak wavelength $\lambda_\text{peak}$ for absorption and gain, using the expression derived in \cite{Myslinski_1997}:
\begin{equation}
\label{eq_abs_nonsat}
\alpha_\text{ns}(\lambda,T)=2k\cdot\alpha_\text{Er}(\lambda,T) \cdot \left(1- \dfrac{\alpha_\text{Er}+g^*_\text{Er}}{2\alpha_\text{Er}+g^*_\text{Er}} \right)(\lambda,T)
\end{equation}
where $\alpha_\text{Er}(\lambda,T)$ and $g^*_\text{Er}(\lambda,T)$ are respectively the absorption and gain spectra at a given temperature (see Fig. \ref{Fig_12_05_spectra}). Our Al:P fiber is a sample of IXF-2CF-EY-PM-12-130 (iXblue): it has a core diameter of $12.6\ \mu$m, a numerical aperture of 0.14, and a clad diameter of $130\ \mu$m. When carefully spliced, the fiber is essentially single-mode, and the overlap between doped core and optical mode is about 90\%. From the peak absorption (see paragraph below) and the typical cross-section of $7.10^{-25}\ $m$^2$ for this type of host \cite{Laroche_2006}, we deduce an erbium concentration [Er]$\approx 2.10^{25}\ m^{-3}$. To saturate the fiber absorption, we used a fiber laser delivering up to 5~W at 1535~nm. We obtained a fraction of paired $\text{Er}^{3+}$ ions $2k=1.6\pm0.5\%$, which is about 6 times lower than the value obtained with standard fibers for that concentration (see $k$ values on Fig.4-9 in \cite{LimPhD}). 

The absorption and gain spectra of this fiber were then measured at 300~K and 77~K (see Fig. \ref{Fig_12_05_spectra}). The usual cut-back method was applied by coupling the broadband spectrum of a halogen lamp in the sample fiber, and measuring the transmission with an optical spectrum analyzer. At $\lambda_\text{peak}=1535$~nm, the small-signal absorption is 46.5~dB/m at 300~K, and 75~dB/m at 77~K, hence a ratio of 1.6 between the Er absorptions at 77 and 300~K. The same way, a ratio of 1.2 was found between the Yb absorptions at 77 and 300~K, for a small signal at 975~nm. This value confirms a previous measurement on Yb-doped fibers at these temperatures \cite{Alimov11}.


The gain spectra were obtained by measuring first the back-propagating fluorescence from a sample pumped in the core at 980 nm through a wavelength division multiplexer (WDM), and correcting from the WDM transmission. The sample was short enough (about 1 cm) to guarantee a full population inversion over its length. This was confirmed by comparing with the spectra obtained for shorter lengths. Then to convert the relative emission spectrum into an absolute gain spectrum, we measured the small-signal gain $g^*_\text{Er}$ at $\lambda_\text{peak}=1535 \ $nm. This was done by injecting a signal seed of about 10$\ \mu$W and measuring the output signal power for various fiber lengths, with the same requirement of full population inversion. The values obtained are $g^*_\text{Er}(\lambda_\text{peak})=45\pm1\ $dB/m at 300~K, and $g^*_\text{Er}(\lambda_\text{peak})=76\pm4\ $dB/m at 77~K. The resulting gain spectra are displayed on Fig. \ref{Fig_12_05_spectra}. 

As expected, according to the Boltzmann statistics, at 77 K the populations of the various sub-levels thermalize to lower energy levels. The number of ions in the lower sublevels of the ground state $^4I_{15/2}$ increases, corresponding to an increase of the absorption close to the zero-phonon line and a decrease at longer wavelengths. Similarly, for an optically pumped ion, the population of the highest sublevels of the excited state decreases, hence reducing the gain at shorter wavelengths.

\subsection{Modeling of a high power EDFA}
The interaction spectra measured above, together with the SBS threshold, are the only parameter that are expected to change notably with the temperature. The laser transition lifetime is another important model parameter, but its value does not vary significantly between 300 and 77~K \cite{Barnes_1991}. Therefore, we can use the results of the previous measurements (Fig. \ref{Fig_12_05_spectra} and \ref{Fig_03_24_comparison_threshold}) to investigate the interest of cooling the Er-doped fiber of an amplifier. We focus our study on the case of a LMA high power EDFA pumped in-band. This type of amplifier is generally interesting to deliver cw or average power in the range of 100~W or higher, with single spatial mode and single frequency, typically for lidar applications or free-space telecommunications \cite{Shi_2014}. Given the high powers involved, eye-safe radiation and low thermal load are required. These requirements can be met by using respectively $1.5\ \mu$m lasers and in-band pumping \cite{Jebali_2014}.

Our simulation program uses the common 4th order Runge-Kutta algorithm to solve the propagation and rate equations along the fiber \cite{Desurvire_EDFA}, with the common approximation of the effective overlap integral between the active region and the propagating mode to suppress the radial dependence \cite{Giles91}. The erbium laser transition is considered as a homogeneously broadened two-level system. This program has been validated by numerous comparisons to experimental results, in various configurations (cw or pulsed regime, core pumped Er or clad pumped Er:Yb fibers...), and yields relative errors lower than 20\% with respect to the output signal and residual pump powers \cite{Canat_2005,Canat_2006}.

The fiber considered here is an LMA fiber with Er:Yb codoping in the same Al:P host as described above, with an absorption of 80~dB/m at $\lambda_\text{peak}$ and 300~K, injected with a cw signal $P_\text{sig}=1\ $W at $\lambda_\text{sig}=1560\ $nm and pumped at $\lambda_\text{pump}=1535\ $nm. The fraction of ions in pairs can be scaled from the value reported above ($2k=1.6\%$ for 46~dB/m), so we consider $2k=3\%$. The effect of the pairs on the efficiency is modeled in a simple way: the powers involved are much higher than the saturation power at both $\lambda_\text{pump}$ and $\lambda_\text{sig}$ (about 3~mW and 20~mW respectively), so the attenuation induced by the ion pairs is essentially constant along the fiber and can be identified to the non-saturable absorption (see eq. (\ref{eq_abs_nonsat})).

The values of the non-saturable absorption calculated at $\lambda_\text{pump}$ and $\lambda_\text{sig}$, at 300 and 77~K, are listed in Table \ref{tab:Fiber_parameters}, and are added to the background loss $l_\text{glass}$ at the corresponding wavelengths. The other model parameters derive from direct measurements or comparisons with results of amplifiers at room temperature with phosphosilicate fibers.

\begin{table}
  \caption{List of the parameters for the in-band pumped Er fiber amplifier model}
  \label{tab:Fiber_parameters} 
\begin{tabular}{|c|c|l|}
  \hline
  Parameter & Value & Comment \\
  \hline
  $\varnothing_\text{core}$ & $20 \ \mu m$ & Core diameter\\
  $\varnothing_\text{clad}$ & $125 \ \mu m$ & Clad diameter\\
  $NA_\text{core}$ & 0.1 & Numerical aperture\\
  $l_\text{glass}$ & 20 dB/km & Background loss \\
  $C_\text{up}$ & $5\ 10^{-24}\ m^3/s $ & Up-conversion factor \\
  $g_B(300 \ \text{K})$ & $10^{-11}$ m/W & SBS gain at 300 K \\
  $g_B(77 \ \text{K})$ & $0.56 \ 10^{-11}$ m/W & SBS gain at 77 K \\
  $\alpha_\text{Er}(1535\ \text{nm})$ & 80 dB/m & Absorption at 300 K\\
  $\alpha_\text{Er}(1535\ \text{nm})$ & 128 dB/m & Absorption at 77 K\\
  $\alpha_\text{ns}(\lambda_\text{pump},300\ \text{K})$ & 20 dB/km & \\
  $\alpha_\text{ns}(\lambda_\text{sig},300\ \text{K})$ & 150 dB/km & \\
  $\alpha_\text{ns}(\lambda_\text{pump},77\ \text{K})$ & 30 dB/km & \\
  $\alpha_\text{ns}(\lambda_\text{sig},77\ \text{K})$ & 16 dB/km & \\
  \hline
\end{tabular}
\end{table}

The model is first used to calculate the longitudinal evolution of $P_\text{pump}(z)$ and $P_\text{sig}(z)$. We then use eq. (\ref{eq_seuilSBS}) to calculate the SBS threshold power along the fiber. This way, we determine for a given pump power the maximum signal output power $P_\text{sig}^{max}$, which is limited either by the pump power (for low pump and corresponding long fiber), or by SBS (for high pump and shorter fiber).

\begin{figure}
	\includegraphics[width=10cm, trim={4cm 2cm 1cm 0}, clip]{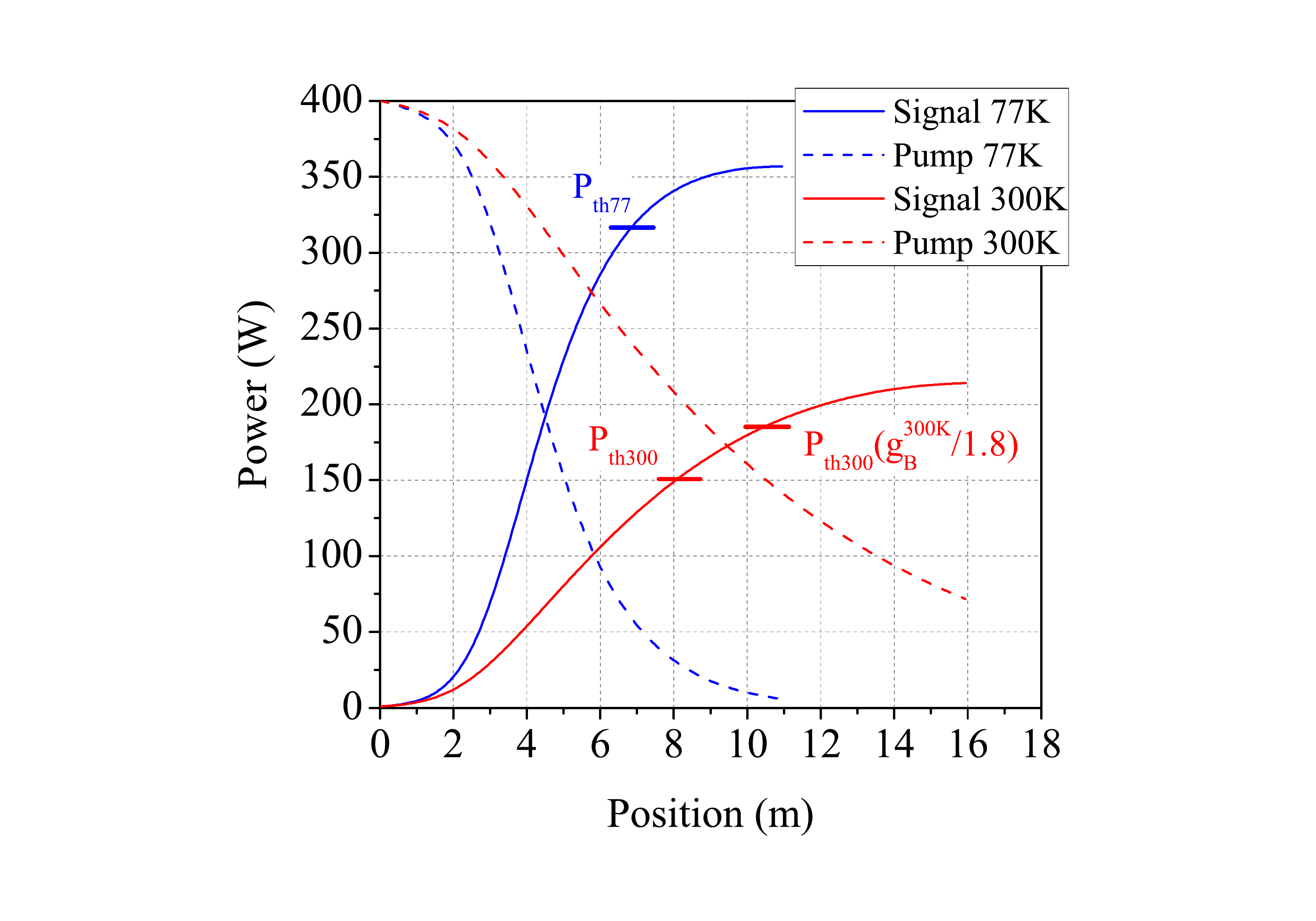}
	\caption{Evolutions of signal (solid) and pump (dashed) powers along the fiber, for an injected pump power of 400 W, at  300~K (red) and 77~K (blue). The SBS thresholds are indicated at 300 and 77~K, as well as the threshold at 300~K if the SBS gain could be reduced to the value of room temperature.}
	\label{Fig_Simu_Exemple_P_vs_z} 
\end{figure}

As an example, for an injected pump power of 400~W, we show on Fig. \ref{Fig_Simu_Exemple_P_vs_z} the different evolutions of $P_\text{pump}(z)$ and $P_\text{sig}(z)$, at 300 and 77~K. The SBS threshold powers are also indicated at the two temperatures, and are obtained from the expression (\ref{eq_interaction_Leff}) of the effective fiber length. From these curves, we extract the maximum signal power as the intersection between the signal $P_\text{sig}(z)$ and the SBS threshold $P_\text{th}(z)$. As expected, the efficiency of the amplifier is higher at low temperature: without taking into account the SBS limitation, the maximum power would be 60\% higher at 77~K than at room temperature (350~W instead of 215~W), for a fiber length 30\% shorter. 

In addition, the SBS gain of a passive silica fiber is 80\% lower at 77~K than at room temperature (see eq. (\ref{eq_ratio_thresholds})). Since the composition of the Er doped silica fiber is almost identical, we assume that the effect of cooling is identical on $g_B$, even if the absolute values may differ. The SBS threshold in the amplifier is then higher at 77~K, with a ratio that depends on the effective lengths of the SBS interaction at the two temperatures. For the example on Fig. \ref{Fig_Simu_Exemple_P_vs_z}, the power is limited by the SBS threshold at both temperatures, and we find $P_\text{sig}^\text{max300}=150\ $W at 300~K and $P_\text{sig}^\text{max77}=315\ $W at 77~K. For an amplifier at room temperature, but with the same value of SBS gain as 77~K (reduced by 80\%), the SBS limited power would increase to 185~W. 

\begin{figure}
	\includegraphics[width=10cm, trim={4cm 1.8cm 1cm 0}, clip]{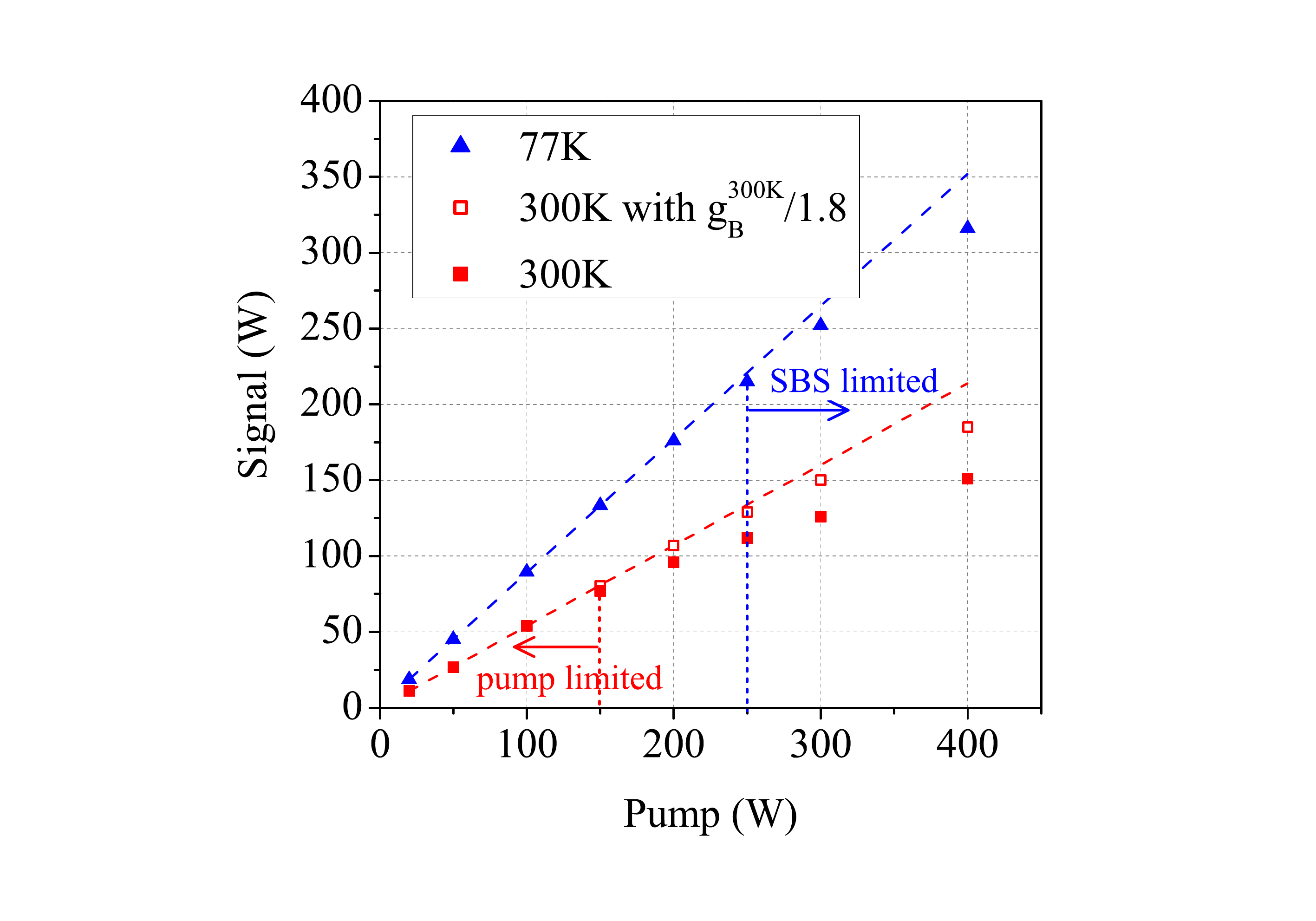}
	\caption{Comparison of the maximum peak signal output as a function of the pump power, for various configurations: at 300~K ($g_B=10^{-11}\ $m/W), at 77~K ($g_B=0.56\ 10^{-11}\ $m/W), and at 300~K with the SBS gain reduced to the same value as 77~K. The dotted lines represent the signal power without SBS limitation.}
	\label{Fig_Simu_Pmax_vs_Pump_300vs77K} 
\end{figure}

The figure \ref{Fig_Simu_Pmax_vs_Pump_300vs77K} shows the variation with $P_\text{pump}$ of the maximum reachable signal power, which is first limited by the pump power, then by the SBS threshold. We compare three configurations: fiber amplifier at 300~K, at 77~K, and at 300~K with the SBS gain reduced by 80\%. In the pump limited regime, i.e. for a pump power lower than 150~W, the optical-optical efficiency (slope of the dotted lines) is about 55\% at 300~K and 90\% at 77~K. The output power is thus superior by 60\% at 77~K, thanks to the sole thermal population distribution. On the other hand in the SBS limited regime (Pump>250~W), the combination of higher acoustic damping and population redistribution at low temperature allow a power increase close to a factor of two.

We also consider the case of an EDFA at 300~K with the SBS gain value of 77~K, in order to compare the distinct influence of the sole SBS gain reduction only and of the fiber cooling. By reducing the SBS gain by 80\%, the output power is improved by a small amount (20\% for a 400~W pump power). This shows that the cold fiber efficiency is the most determining contribution in the improvement of the output power. In addition, let us note that the experimental techniques allowing the SBS gain reduction (high doping, fiber strain, signal broadening, etc) could also be applied at low temperature, increasing again the maximum output power.

As argued in the beginning, we find that cooling the fiber can bring together the two benefits of higher SBS threshold and higher amplifier efficiency. For all the points calculated, the amplified spontaneous emission (ASE) at the fiber input and output was negligible. 

\section{Argument about the laser generated heat}

One could be concerned about the heat generated by this high power setup in the fiber core, which could increase the temperature. We show that there is no such limitation, thanks to the low quantum defect between pump and signal: most of the pump energy is converted into signal. We assume here that the main contribution to the heat would come from the non-radiative decay of the pump laser. Considering the same example presented previously (see Fig. \ref{Fig_Simu_Exemple_P_vs_z}), we calculate the temperature increment caused by the fraction of pump that is not converted into signal. At 77~K, the highest rate of pump absorption is about 90~W/m (between 3 and 5~m on Fig. \ref{Fig_Simu_Exemple_P_vs_z}). The quantum defect between pump and signal radiations is given by $1-\lambda_\text{pump}/\lambda_\text{signal} \approx 2\%$, so the power that is converted into heat is lower than 2~W/m. 

To convert this heat load into a quantitative temperature increase in the fiber core, the fiber is modeled with a finite element method. We consider that the fiber is cooled in a standard cryogenic setup, like the widespread Stirling machines used for infrared detectors. The system can be described as the succession of layers. The Er doped silica core is surrounded by a silica clad, with basically the same thermal properties. A cladding of acrylate polymer protects mechanically the silica fiber. Then the fiber is glued with epoxy between two copper plates. The successive stages of thermal contacts are supposed faultless. The thickness of each layer is reported in Table \ref{tab:Fiber_layers}, together with the thermal conductivity at 77~K.

\begin{table}
  \caption{List of the layers used to calculate the temperature increase caused in the fiber core by the pump laser.}
  \label{tab:Fiber_layers} 
\begin{tabular}{|c|c|c|}
  \hline
  Layer & Thickness & Thermal \\
  & & conductivity (77 K) \\
  \hline
  Silica core & $20 \ \mu$m & 0.6~W/m/K\\
  Silica clad & $60 \ \mu$m & 0.6~W/m/K\\
  Polymer coating & $115 \ \mu$m & 0.1~W/m/K \\
  Epoxy glue (square) & $200 \ \mu$m & 1~W/m/K \\
  Copper base \& top  & 5 mm & 500~W/m/K \\
  \hline
\end{tabular}
\end{table}

These parameters are used to calculate the temperature across the setup, for the region that experiences the highest thermal load (2~W/m, i.e. 6.37e9~W/m$^3$ in the core). To simplify, we assume that the fiber coils are not in contact with each other. 


\begin{figure}
	\includegraphics[width=6cm, trim={0cm 0 -0cm 0}, clip]{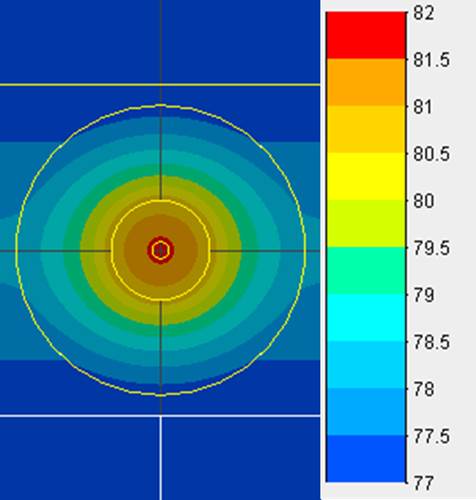}
	\caption{Temperature distribution across the section of the fiber, calculated with the parameters listed above.}
	\label{Fig_Temp_gradient_cross_section} 
\end{figure}

As can be seen on Fig. \ref{Fig_Temp_gradient_cross_section}, the temperature in the fiber core increases by about 5 K in this section of highest absorption, for a pump power of 400~W. This calculation shows that despite the high pump power coupled into the fiber clad, the temperature does not increase significantly compared to the variation that would be required to affect the SBS threshold and the laser efficiency (see Fig. \ref{Fig_03_23_linewidth_empirical_def}).

\section{Conclusion}
Taking advantage of the well-referenced effects of cryogenic temperatures on both phonon lifetime and on laser efficiency in silica fibers, we present the first study of their combined advantage for the performance of a single-mode and single-frequency Er-doped fiber laser amplifier. The SBS threshold power is found 80\% higher at 77~K than at 300~K. At the same temperatures, we report the measurement of the absorption and gain spectra of the $^5I_{13/2}-^5I_{15/2}$ laser transition of erbium, in a Er:Yb codoped alumino-phosphosilicate host.

These data are used in a numerical simulation of the signal amplification at 1560~nm in the same type of fiber, clad pumped at 1535~nm, with a pump power up to 400~W. In the pump limited regime, the amplifier at low temperature is 60\% more efficient. In the SBS limited regime, given the higher SBS threshold power at 77~K, the amplifier can reach a power more than twice higher than at room temperature. Let us note that the cryogenic cooling can be combined with a number of other methods of SBS rejection, such as fibers with large mode area, low Brillouin gain (anti-guiding profile for acoustic waves), strain gradient, or signal frequency broadening... 

This paper demonstrates the interest of cooling the fiber, so the experiment should be tested in a configuration similar to that suggested here. In a longer term perspective, our work also illustrates the interest of understanding and using the physical properties of non-usual amorphous hosts for improving the performance of rare-earth doped fibers, notably in laser amplifiers. Historically, silica fibers have been used for EDFA because of their immediate compatibility (fabrication, splicing...) with the low absorption fibers used for long-distance telecommunications. Since this technological leap, many other glasses or doping methods were investigated \cite{Hertling_1998, Shi2010, Lambin_2013, Ballato2013, Savelii_2017, Dragic_2018} to reduce the SBS gain or improve the laser amplification efficiency, but few experiments have demonstrated a positive combination on these two parameters.


\section*{Acknowledgment}

We would like to thank Marie Foret from Laboratoire Charles Coulomb, Université Montpellier, for helpful discussions. We are also thankful to Agnès Dolfi-Bouteyre and François Gustave from ONERA, and Jean-Louis Le Gouët from Laboratoire Aimé Cotton (CNRS, UPR 3321) for careful reading.

\ifCLASSOPTIONcaptionsoff
  \newpage
\fi

\begin{IEEEbiography}{Julien Le Gouët}received the Graduate degree from the Ecole Supérieure d’Optique (Orsay, France) in 2004, and the PhD degree on cold atom gravimetry at the Observatoire de Paris and Université Paris XI in 2008. After a post-doctoral stay in the Optical and Quantum Communications group of the Research Lab of Electronics at MIT, Massachusetts, he joined the Laser Sources and LIDAR Systems unit of the Office National des Etudes et Recherches Aérospatiales (ONERA) in 2010. His main research interests include rare-earth ions spectroscopy in optical fibers, fiber laser amplifiers for coherent lidars, and high power laser sources.
\end{IEEEbiography}

\begin{IEEEbiographynophoto}{Jérémy Oudin}
Biography text here.
\end{IEEEbiographynophoto}


\begin{IEEEbiographynophoto}{Philippe Perrault}
Biography text here.
\end{IEEEbiographynophoto}

\begin{IEEEbiographynophoto}{Alaeddine Abbes}
Alaeddine Abbes received the M.Sc. degree in Optoelectronics and Microwave Engineering in 2017, from the University of Montpellier. He is currently working toward the Ph.D. degree on the development of a compact, low-cost, highly tunable, highly coherent and powerful terahertz (THz) sources based on photomixing using VeCSEL (Vertical-external-Cavity Surface-Emitting-Laser) technology from the Institut d'Électronique et des Systèmes, Montpellier, France. He current research interests include THz generation and detection, THz optoelectronic generation using photoconductors and photodiodes, THz antenna, and semiconductor lasers.
\end{IEEEbiographynophoto}

\begin{IEEEbiographynophoto}{Alice Odier}
Biography text here.
\end{IEEEbiographynophoto}

\begin{IEEEbiographynophoto}{Alizée Dubois}
Biography text here.
\end{IEEEbiographynophoto}




\begin{thebibliography}{1}

\bibitem{Dolfi_2009}
A. Dolfi-Bouteyre, G. Canat, M. Valla, B. Augère, C. Besson, D. Goular, L. Lombard, J.-P. Cariou, A. Durecu, D. Fleury, L. Bricteux, S. Brousmiche, S. Lugan, and B. Macq, “Pulsed 1.5$\mu$m lidar for axial aircraft wake vortex detection based on high brightness large-core fiber amplifier,” IEEE J. Sel. Top. Quantum Electron. \textbf{15}, pp. 441--450, 2009. 

\bibitem{Creeden_2016}
D. Creeden, H. Pretorius, J. Limongelli, and S. D. Setzler, "Single frequency 1560 nm Er:Yb fiber amplifier with 207 W output power and 50.5\% slope efficiency," Proc. SPIE 9728, 97282, 2016.

\bibitem{Agrawal_book}
G. P. Agrawal, \textit{Nonlinear Fiber Optics}, 3rd ed., Academic press, 2001. 

\bibitem{Ahmad2017}
R. Ahmad,  M. F. Yan, J. W. Nicholson, K. S. Abedin, P. S. Westbrook, C. Headley, P. W. Wisk, E. M. Monberg and D. J. DiGiovanni, "Polarization-maintaining, large-effective-area, higher-order-mode fiber," Opt. Lett. \textbf{42}, 13, pp. 2591--2594, 2017.

\bibitem{Alimov11}
O. K. Alimov, T. T. Basiev, V. A. Konushkin, A. G. Papashvili, A. Y. Karasik and L. Henry, "Investigations of Yb-doped optical fiber using selective laser excitation", Applied Physics B \textbf{104}, 2011.

\bibitem{Anderson55}
O. L. Anderson and H. E. Bömmel, "Ultrasonic Absorption in Fused Silica at Low Temperatures and High Frequencies", Journal of the American Ceramic Society \textbf{38}, pp. 125--131, 1955.

\bibitem{Ballato2013}
J. Ballato and P. Dragic, "Rethinking Optical Fiber: New Demands, Old Glasses", Journal of the American Ceramic Society \textbf{96}, pp. 2675--2692, 2013.

\bibitem{Barnes_1991}
W. L. Barnes, R. I. Laming, E. J. Tarbox and P. R. Morkel, "Absorption and emission cross section of Er$^{3+}$ doped silica fibers," in IEEE Journal of Quantum Electronics \textbf{27}, pp. 1004-1010, 1991.


\bibitem{Boyd_ONL}
R. W. Boyd, \textit{Nonlinear Optics}, second ed., Academic press, 2003.

\bibitem{Boyd_1990}
R. W. Boyd, K. Rzazewski, and P. Narum, "Noise initiation of stimulated Brillouin scattering," Phys. Rev. A \textbf{42}, pp. 5514-5521, 1990. 

\bibitem{Bubnov2009}
M. M. Bubnov, A. N. Gur'yanov, K. V. Zotov, L. D. Iskhakova, S. V. Lavrishchev, D. S. Lipatov, M. E. Likhachev, A. A. Rybaltovsky, V. F. Khopin, M. V. Yashkov, and E. M. Dianov, "Optical properties of fibres with aluminophosphosilicate glass cores", Quantum Electronics \textbf{39}, p. 857, 2009.


\bibitem{Canat_2005}
G. Canat, J.-C. Mollier, Y. Jaouën, and B. Dussardier, "Evidence of thermal effects in a high-power Er3+-Yb3+ fiber laser," Opt. Lett. \textbf{30}, pp. 3030--3032, 2005.

\bibitem{Canat_2006}
G. Canat, Y. Jaouën, and J.-C. Mollier, "Performance and limitations
of high brightness Er-Yb fiber sources," Comptes Rendus Physique \textbf{7}, pp. 177--186, 2006.

\bibitem{Courtens07}
E. Courtens, M. Foret, B. Rufflé, and R. Vacher, "Elasticity and anelasticity of oxide glasses," Eur. J. Glass Sci. and Technol. B \textbf{48}, pp. 9--18, 2007.

\bibitem{Desurvire_1990}
E. Desurvire, J. Zyskind, and J. R. Simpson, "Spectral gain hole-burning at 1.53 $\mu$m in erbium-doped fiber amplifiers," IEEE Photon. Technol. Lett. \textbf{2}, pp. 246--248, 1990.

\bibitem{Desurvire_EDFA}
E. Desurvire, Erbium-Doped Fiber Amplifiers: Principles and Applications, Wiley ed., 1994.

\bibitem{DiGiovanni1989}
D. DiGiovanni, J. MacChesney, and T. Kometani, "Structure and properties of silica containing aluminum and phosphorus near the AlPO$_4$ join," Journal of Non-Crystalline Solids \textbf{113}, pp. 58--64, 1989.

\bibitem{Dragic_2018}
P. D. Dragic, M. Cavillon, and J. Ballato, "Materials for optical fiber lasers: A review," Applied Physics Reviews \textbf{5}, 041301, 2018.

\bibitem{Dubinski09}
N. Ter-Gabrielyan, M. Dubinskii, G. Newburgh, A. Michael, and L. D. Merkle, "Temperature dependence of a diode-pumped cryogenic Er:YAG laser," Opt. Express \textbf{17}, pp. 7159--7169, 2009.

\bibitem{EPFL02}
L. Thévenaz, A. Fellay, M. Facchini, W. Scandale, M. Nikles, and P. Robert, "Brillouin optical fiber sensor for cryogenic thermometry," in \textit{Proceedings of SPIE}, Vol. 4694, 2002.

\bibitem{Giles91}
C. Giles and E. Desurvire, "Modeling erbium doped fiber amplifiers," J. Lightwave Technol. \textbf{9}, pp. 271--283, 1991.

\bibitem{Hertling_1998}
J. Hertling, S. Baessler, S. Rau, G. Kasper, and S. Hunklinger, "Internal friction and hypersonic velocity in vitreous germania under high pressure," Journal of Non-Crystalline Solids 226, pp. 129 -- 137, 1998.

\bibitem{Imai_1993}
Y. Imai and N. Shimada, "Dependence of stimulated Brillouin scattering on temperature distribution in polarization-maintaining fibers," IEEE Photonics Technology Letters 5, pp. 1335--1337, 1993.

\bibitem{Jebali_2014}
M. A. Jebali, J.-N. Maran, and S. LaRochelle, "264 W output power at 1585 nm in Er-Yb codoped fiber laser using in-band pumping," Opt. Lett. \textbf{39}, pp. 3974--3977, 2014.

\bibitem{Jelger09}
P. Jelger, K. Seger, V. Pasiskevicius, and F. Laurell, "Highly efficient temporally stable narrow linewidth cryogenically cooled Yb-fiber laser," Opt. Express \textbf{17}, 2009.

\bibitem{Kagi_91}
N. Kagi, A. Oyobe, and K. Nakamura, "Temperature dependence of the gain in erbium doped fibers," J. Lightwave Technol. \textbf{9}, pp. 261--265, 1991.

\bibitem{Kiritchenko2015}
N. V. Kiritchenko, L. V. Kotov, M. A. Melkumov, M. E. Likhachev, M. M. Bubnov, M. V. Yashkov, A. Y. Laptev, and A. N. Guryanov, "Effect of ytterbium co-doping on erbium clustering in silica-doped glass," Laser Physics \textbf{25}, 025102, 2015.

\bibitem{Kobyakov10}
A. Kobyakov, M. Sauer, and D. Chowdhury, "Stimulated Brillouin scattering in optical fibers," Advances in Optics and Photonics \textbf{2}, pp. 1--59, 2010.

\bibitem{Lambin_2013}
V. Lambin-Iezzi, S. Loranger, M. Saad, R. Kashyap, "Stimulated Brillouin scattering in SM ZBLAN fiber", J. Non-Cryst. Solids \textbf{359}, pp. 65–68, 2013.

\bibitem{Laroche_2006}
M. Laroche, S. Girard, J. K. Sahu, W. A. Clarkson, and J. Nilsson, "Accurate efficiency evaluation of energy-transfer processes in phosphosilicate Er3+-Yb3+-codoped fibers," J. Opt. Soc. Am. B \textbf{23}, pp. 195--202, 2006.

\bibitem{Lefloch01}
S. Le Floch, F. Riou, and P. Cambon, "Experimental and theoretical study of the Brillouin linewidth and frequency at low temperature in standard single-mode optical fibres," J. Opt. A: Pure Appl. Opt. \textbf{3}, 2001.

\bibitem{Lefloch03}
S. Le Floch and P. Cambon, "Study of Brillouin gain spectrum in standard single-mode optical fiber at low temperatures (1.4-370 K) and high hydrostatic pressures (1-250 bars)," Opt. Commun. \textbf{219}, pp. 395--410, 2003.

\bibitem{Li2007}
M.-J. Li, X. Chen, J. Wang, S. Gray, A. Liu, J. A. Demeritt, A. B. Ruffin, A. M. Crowley, D. T. Walton, and L. A. Zenteno, "Al/Ge co-doped large mode area fiber with high SBS threshold," Opt. Express \textbf{15}, pp. 8290--8299, 2007.

\bibitem{Lim2012}
E.-L. Lim, S.-U. Alam, and D. J. Richardson, "Optimizing the pumping configuration for the power scaling of in-band pumped erbium doped fiber amplifiers," Opt. Express \textbf{20}, pp. 13886--13895, 2012.

\bibitem{LimPhD}
E.-L. Lim, \textit{Pump conditioning and optimisation for erbium doped fibre applications}, Ph.D. thesis, University of Southampton, 2012.

\bibitem{Millar_90}
C. A. Millar, T. J. Whitley, and S. C. Fleming, "Thermal properties of an erbium-doped fiber amplifier," Proc. Inst. Elect. Eng. \textbf{137}, pp. 155--162, 1990.

\bibitem{Myslinski_1997}
P. Myslinski, D. Nguyen, and J. Chrostowski, "Effects of concentration on the performance of erbium-doped fiber amplifiers," Journal of Lightwave Technology \textbf{15}, pp. 112--120, 1997.

\bibitem{MIT08}
S. Mahar, J. Geng, J. Schultz, J. Minervini, S. Jiang, P. Titus, M. Takayasu, C.-Y. Gung, W. Tian, and A. Chavez-Pirson, "Real-time simultaneous temperature and strain measurements
at cryogenic temperatures in an optical fiber," in \textit{SPIE}, Vol. 7087, 2008.


\bibitem{Pine69}
A. S. Pine, "Brillouin scattering study of acoustic attenuation in fused quartz," Physical Review
\textbf{185}, p. 1187, 1969.

\bibitem{Savelii_2017}
I. Savelii, L. Bigot, B. Capoen, C. Gonnet, C. Chanéac, E. Burova, A. Pastouret, H. El-Hamzaoui and M. Bouazaoui, "Benefit of Rare-Earth "Smart Doping" and Material Nanostructuring for the Next Generation of Er-Doped Fibers," Nanoscale Research Letters \textbf{12}, 2017.

\bibitem{Shi2010}
W. Shi, E. B. Petersen, Z. Yao, D. T. Nguyen, J. Zong, M. A. Stephen, A. Chavez-Pirson, and N. Peyghambarian, "Kilowatt-level stimulated-Brillouin-scattering-threshold monolithic transform-
limited 100~ns pulsed fiber laser at 1530 nm," Opt. Lett. \textbf{35}, pp. 2418--2420, 2010.

\bibitem{Shi_2014}
W. Shi, Q. Fang, X. Zhu, R. A. Norwood, and N. Peyghambarian, "Fiber lasers and their applications," Appl. Opt. \textbf{53}, pp. 6554--6568, 2014.

\bibitem{Steinborn13}
R. Steinborn, A. Koglbauer, P. Bachor, T. Diehl, D. Kolbe, M. Stappel, and J. Walz, "A continuous wave 10 W cryogenic fiber amplifier at 1015 nm and frequency quadrupling to 254 nm," Opt. Express
\textbf{21}, 2013.

\bibitem{Tittel_2015}
L. Veissier, M. Falamarzi, T. Lutz, E. Saglamyurek, C. W. Thiel, R. L. Cone, and W. Tittel, "Optical decoherence and spectral diffusion in an erbium-doped silica glass fiber featuring long-lived spin sublevels," Phys. Rev. B \textbf{94}, 195138, 2016.

\bibitem{Unger2007}
S. Unger, A. Schwuchow, J. Dellith, and J. Kirchhof, "Codoped materials for high power fiber lasers: discussion behaviour and optical properties," in \textit{Proceedings of SPIE}, Vol. 6469, 2007.

\bibitem{Vienne_1998}
G. G. Vienne, J. E. Caplen, L. Dong, J. D. Minelly, J. Nilsson, and D. N. Payne, "Fabrication and Characterization of Yb3+ : Er3+ Phosphosilicate Fibers for Lasers," Journal of Lightwave Technology \textbf{16}, pp. 1990--2001, 1998.

\bibitem{Vienne1996}
G. G. Vienne, W. Brocklesby, R. Brown, Z. Chen, J. Minelly, J. Roman, and D. Payne, "Role of aluminum in ytterbium:erbium codoped phosphoaluminosilicate optical fibers," Optical Fiber Technology \textbf{2}, pp. 387--393, 1996.

\bibitem{Yao2011}
T. F. Yao, R. Steinborn, A. S. Webb, J. K. Sahu, and J. Nilsson, "Cryogenically cooled erbium-doped fiber lasers," in \textit{International Summer Session: Lasers and Their Applications}, p. Th12, 2011.

\bibitem{Yoshizawa1993}
N. Yoshizawa and T. Imai, "Stimulated Brillouin scattering suppression by means of applying strain distribution to fiber with cabling," J. Lightwave Technol. \textbf{11}, pp. 1518--1522, 1993.


\end{thebibliography}
\end{document}